\documentclass[twocolumn,trackchanges]{aastex631}

\newcommand{\sfr}{$\rm M_{\odot}~yr^{-1}$}
\newcommand{\ergs}{erg s$^{-1}$}
\newcommand{\kms}{km s$^{-1}$}
\newcommand{\msun}{$\rm M_{\odot}$}
\usepackage{amsmath}

\begin{document}

\title{The resolved star formation law in NGC 7469 from JWST, ALMA and VLA}

\correspondingauthor{Maria Vittoria Zanchettin}
\email{mazanch@sissa.it}

\author[0000-0001-7883-496X]{Maria Vittoria Zanchettin}
\affiliation{SISSA, Via Bonomea 265, I-34136 Trieste, Italy}
\affiliation{INAF Osservatorio Astronomico di Trieste, via G.B. Tiepolo 11, 34143 Trieste, Italy}

\author[0000-0002-0375-8330]{Marcella Massardi}
\affiliation{INAF, Istituto di Radioastronomia, Italian ARC, Via Piero Gobetti 101, I-40129 Bologna, Italy}

\author[0000-0003-4751-7421]{Francesco Salvestrini}
\affiliation{INAF Osservatorio Astronomico di Trieste, via G.B. Tiepolo 11, 34143 Trieste, Italy}

\author[0000-0002-4314-021X]{Manuela Bischetti}
\affiliation{Dipartimento di Fisica, Università di Trieste, Sezione di Astronomia, Via G.B. Tiepolo 11, I-34131 Trieste, Italy}
\affiliation{INAF Osservatorio Astronomico di Trieste, via G.B. Tiepolo 11, 34143 Trieste, Italy}

\author[0000-0002-4227-6035]{Chiara Feruglio}
\affiliation{INAF Osservatorio Astronomico di Trieste, via G.B. Tiepolo 11, 34143 Trieste, Italy}
\affiliation{IFPU - Institute for fundamental physics of the Universe, Via Beirut 2, 34014 Trieste, Italy}

\author[0000-0002-4031-4157]{Fabrizio Fiore}
\affiliation{INAF Osservatorio Astronomico di Trieste, via G.B. Tiepolo 11, 34143 Trieste, Italy}
\affiliation{IFPU - Institute for fundamental physics of the Universe, Via Beirut 2, 34014 Trieste, Italy}

\author[0000-0002-4882-1735]{Andrea Lapi}
\affiliation{SISSA, Via Bonomea 265, I-34136 Trieste, Italy}
\affiliation{IFPU - Institute for fundamental physics of the Universe, Via Beirut 2, 34014 Trieste, Italy}
\affiliation{INAF, Istituto di Radioastronomia, Italian ARC, Via Piero Gobetti 101, I-40129 Bologna, Italy}
\affiliation{INFN, Sezione di Trieste, via Valerio 2, Trieste I-34127, Italy}



\begin{abstract}
We investigate the star formation process within the central 3.3 kpc region of the nearby luminous infrared Seyfert NGC 7469, probing scales ranging from 88 to 330 pc. 
We combine JWST/MIRI imaging with the F770W filter, with CO(2-1) and the underlying 1.3 mm dust continuum data from ALMA, along with VLA radio continuum observations at 22 GHz. 
NGC 7469 hosts a starburst ring which dominates the overall star formation activity. We estimate a global star formation rate SFR $\sim 11.5$ \sfr\ from the radio at 22 GHz, and a cold molecular gas mass $\rm M(H2) \sim 6.4 \times 10^9 M_{\sun}$ from the CO(2-1) emission. 
We find that the 1.3 mm map shows a morphology remarkably similar to those traced by the 22 GHz and the 7.7 \micron\ polycyclic aromatic hydrocarbon (PAH) emission observed with JWST. The three tracers reproduce the morphology of the starburst ring with good agreement.
We further investigate the correlations between the PAHs, the star formation rate and the cold molecular gas.
We find a stronger correlation of the PAHs with the star formation than with the CO, with steeper correlations within the starburst ring ($n > 2$) than in the outer region ($n < 1$). 
We derive the correlation between the star formation rate and the cold molecular gas mass surface densities, the Kennicutt-Schmidt star formation law. Comparisons with other galaxy populations, including starburst galaxies and active galactic nuclei, highlighted that NGC 7469 exhibits an intermediate behavior to the Kennicutt-Schmidt relations found for these galaxy populations. 
\end{abstract}



\section{Introduction} \label{sec:intro}

\citet{schmidt1959,schmidt1963} proposed the existence of empirical scaling relations between the properties of the interstellar gas and the star formation rate (SFR) in the same environment. Later, \citet{kennicutt1998a,kennicutt1998b} confirmed the correlation between global measurements of galaxy gas mass and SFR in nearby galaxies. Consequently, this correlation has become known as the Kennicutt-Schmidt (K-S) star formation law.

Spatially unresolved studies that target molecular gas and SFR help to derive global disk properties and are necessary to test the redshift dependence of the gas–star formation relation \citep{genzel2010, schruba2011}.
However, to understand the physical mechanisms behind the star formation (SF) process, spatially resolved measurements in nearby targets are necessary. These observations allow us to study how the SF properties, such as the timescales for gas consumption and the SF efficiency, vary across the galaxy environments \citep{onodera2010,bigiel2011, leory2013, ford2013, kreckel2018, dey2019,pessa2022,sun2023}.
The tracers most commonly used to quantify the SFR are the most commonly observed optical spectral lines (e.g. $\rm H_{\alpha}$ and [O II]), the global radio continuum, mid- and far- infrared (MIR and FIR), and ultraviolet (UV) emission. Such SFR indicators refer to various wavelengths that trace stellar populations at different evolutionary stages \citep[see][for a review]{kennicutt2012}.
Optical lines and UV light are emitted predominantly by young stars and are sensitive to SF on $\sim$10 Myr and $\sim$10–100 Myr timescales respectively \citep[][]{hao2011,murphy2011}. However, accurate SFR estimates from UV data are affected by large uncertainties due to the dust extinction correction \citep{mahajan2019}. The thermal free-free and non-thermal synchrotron radio emission at low frequency (1-10 GHz) are a measure of past formation of massive stars, resulting in a SF time sensitivity of about 150-300 Myr \citep{klein1981, gioia1982, tabatabaei2017, arango-toro2023}.
The MIR (3–25 \micron) emission of star forming galaxies is dominated by the polycyclic aromatic hydrocarbon (PAH) features both at low and high redshift \citep{puget1989,genzel1998,armus2007,valiante2007,farrah2008,veilleux2009,Riechers2014}. In particular PAHs has been proposed as a good indicator of SFR over timescales of up to a few hundred million years \citep{peeters2004,shipley2016,xie2019}.
Moreover a connection between PAHs and the properties of the cold molecular gas, traced by the CO emission, has been proposed by several studies \citep{cortzen2019,alonso-herrero2020}.
Taking advantage of interferometric mm and cm, and JWST data of nearby galaxies, it is possible to perform spatially resolved studies of the SF process in galaxies, down to hundred-parsec scales and to investigate the correlations between the PAHs, the SFR surface density ($\rm \Sigma_{SFR}$) and the mass surface density of molecular gas ($\rm \Sigma_{H2}$), which is the direct fuel of the formation of stars.

In this paper, we study the SF process in the inner ($\sim$3 kpc) region of NGC 7469, a nearby (z = 0.01627) luminous infrared galaxy ($\rm L_{IR, 8-1000 \rm \mu m} = 10^{11.6} L_{\odot}$), and type-1.5 Seyfert \citep{landt2008} with a bolometric AGN luminosity of $\rm L_{bol,AGN}=3.5\times 10^{44}$ \ergs \citep{gruppioni2016}. 
From VLA 8.4 GHz observations at 0.3 arcsec resolution, \citet{perez-torres2009} estimated the nuclear radio emission $\rm L_{8.4~GHz} \sim  10^{36.96} $ \ergs\ . NGC 7469 may host a radio core–jet structure as resolved on 0.3 arcsec scale by the VLA \citep{lodsdale2003,alberdi2006,orientiprieto2010}, and no evidence of a radio jet–ISM interaction has yet been reported \citep{xuwang2022}.
NGC 7469 is known to host a prominent starburst (SB) ring with an outer radius of 900 pc, which dominates the global SFR of $\sim 20-50$ \sfr \citep{genzel1995,soifer2003,gruppioni2016,song2021}.
A bar-like feature connects the SB ring to a circumnuclear disk (CND) located in the inner 300 pc \citep[see][and references therein]{davies2004,izumi2015,izumi2020}.
The molecular gas mass derived from CO(1-0) within the inner $\sim$1.2 kpc region is $\rm M(H2) = 2.7 \times 10^9 M_{\odot}$ \citep{davies2004,izumi2020}.
To investigate the SF process inside and outside the SB ring of NGC 7469, we combine James Webb Space Telescope (JWST) imaging with the F770W filter of the Mid-InfraRed Instrument (MIRI), with CO(2-1) and the underlying dust continuum (224.2-243.7 GHz) data obtained from the Atacama Large Millimeter/submillimeter Array (ALMA), along with the Karl G. Jansky Very Large Array (VLA) radio continuum imaging at 22 GHz, resolving physical scales in the range 88 pc (MIRI) to 330 pc (VLA). 

This Paper is organized as follows. In Section \ref{sec:obs}, we describe the observations and data reduction. Section \ref{sec:analysis} provides details on the data analysis and quantity derivations. Section \ref{sec:discussion} describes our results. The adopted luminosity distance of the galaxy is 72.1 Mpc and the corresponding scale is 0.338 kpc/arcsec. 

\section{Observations and Data Reduction} \label{sec:obs}

\begin{figure*}[ht]
  \resizebox{\hsize}{!}{ \includegraphics{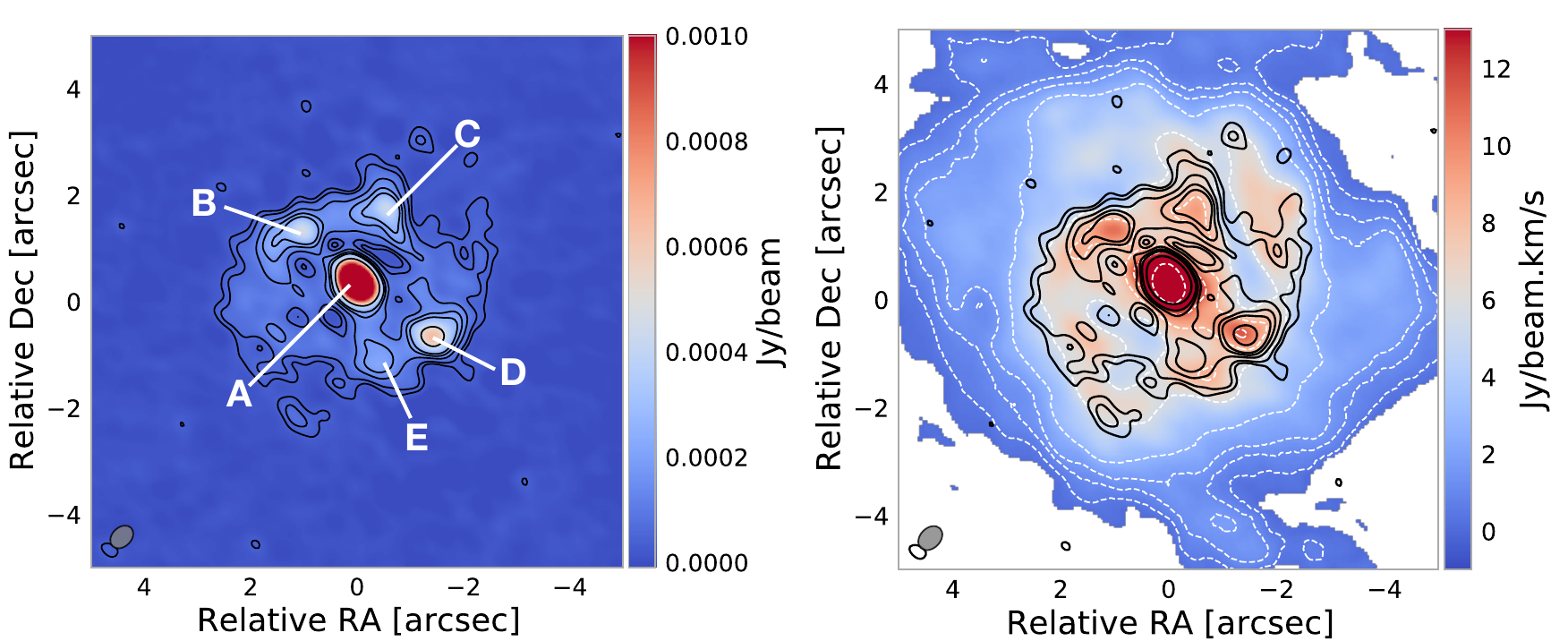}}
 \caption{Left panel: 1.3 mm continuum map. Labels from A to E indicate the five clumps of emission. The position of A corresponds to the AGN position \citep[23:03:15.617 +08:52:26.00 J2000,][]{izumi2020}. Contours are drawn at (3, 5, 10, 15, 20, 40)$\rm \sigma$, $\rm \sigma = 0.0121$ mJy/beam. The clean beam (0.492 arcsec $\times$ 0.360 arcsec, PA=46 deg) is indicated by the gray ellipse. Right panel: velocity integrated CO(2-1) (moment-0) map. White contours are drawn at (3, 5, 10, 15, 20, 30)$\rm \sigma$, $\rm \sigma = 0.56$ Jy/beam.km/s. The clean beam (0.508 $\times$ 0.374 arcsec, PA=45 deg) is indicated by the gray ellipse. Regions with emission below 1$\rm \sigma$ are blanked out.}
  \label{fig:alma}
\end{figure*}

We analyzed band 6 ALMA 12 m observations (program ID 2017.1.00078.S, PI Izumi Takuma) of NGC 7469 of the CO(2-1) line (observed frequency 226.85 GHz) and the underlying continuum (224.2-243.7 GHz). 
NGC 7469 was observed in September 2018 by using the 12m array for a total integration time of 1.48 h in the configuration C43-4 with 44 antennas, proving an angular resolution of about 0.4 arcsec and a largest angular scale of 4.8 arcsec.
Visibilities were calibrated by the NRAO Observatory, using the standard pipeline and default calibrators. Imaging was performed with the Common Astronomy Software Applications \citep[CASA;][]{mcmullin}, version 5.1.1-5.
We created a continuum map by averaging the visibilities in the four spectral windows (spws), excluding the spectral range covered by the CO(2-1) emission line. We produced continuum-subtracted datacubes by fitting and subtracting a first-order polynomial from the visibilities of the spw covering CO(2-1). Imaging was performed with task tclean using the \texttt{hogbom} algorithm, a threshold of three times the root mean square (rms) noise, a briggs weighting scheme with robust parameter $b =$ 0.5 for the continuum map and a natural weighting scheme for the CO(2-1) data cube in order to increase the map sensitivity. 
The resulting synthesized beam for the continuum map is 0.492 arcsec $\times$ 0.360 arcsec (corresponding to 164 pc $\times$ 120 pc), with an rms of 0.0121 mJy/beam. The CO(2-1) data cube has a beam size of 0.508 arcsec $\times$ 0.374 arcsec and a rms noise of 0.518 mJy/beam for a channel width of 5 \kms.
This translates into a mass surface density sensitivity of 7.8 \msun /beam for a channel width of 5 \kms, computed using equation \ref{eq:MH2}.
Figure \ref{fig:alma} shows (left panel) the millimetre continuum map and (right panel)
the continuum-subtracted velocity-integrated map of the CO(2-1) emission.

We analysed archival VLA observations of NGC 7469 in band K (project code 20A-158, PI Krista L. Smith). The observations were carried out in April 2020 for a total integration time of 58.9 h in the frequency range 18-26 GHz. We used the standard calibrated visibilites provided by the VLA data archive l\footnote{\href{(https://science.nrao.edu/facilities/vla/archive}{(https://science.nrao.edu/facilities/vla/archive}} and we used CASA 5.1.1-5 software to generate the map of K band continuum. We averaged the 62 spw and we produced a clean map using the \texttt{tclean} task and a briggs weighting scheme with robust parameter $b =$ 0.5. We used the \texttt{hogbom}
cleaning algorithm with a detection threshold of 3 times the rms sensitivity.
The final continuum map has an rms of 23.2 $\rm \mu Jy/beam $ and a clean beam of 1.029 arcsec $\times$ 0.967 arcsec (i.e. 342 pc $\times$ 322 pc).

JWST/MIRI \citep{bouchet2015, rieke2015} imaging of NGC 7469 was obtained with the F770W (7.7 $\mathrm{\mu m}$) filter with both the BRIGHTSKY and SUB128 subarray modes (program ID 1328, PI Armus Lee). In this paper we focus the analysis on the SUB128 subarray data, where the nucleus remained unsaturated, taken on 2022 July 2 with an exposure time of 48 s. The field of view of 14.8 arcsec $\times$ 16.4 arcsec and the pixel size is 0.11 arcsec. The calibrated JWST/MIRI 7.7 \micron\ map has a sensitivity of 12 $\rm MJy \, sr^{-1}\, pixel^{-1}$.
We used the calibrated images provided by the STSCI Observatory, downloaded through the MAST portal\footnote{\href{https://mast.stsci.edu/portal/Mashup/Clients/Mast/Portal.html}{https://mast.stsci.edu/portal/Mashup/Clients/Mast/Portal.html}}.

\begin{figure*}[ht]
  \resizebox{\hsize}{!}{ \includegraphics{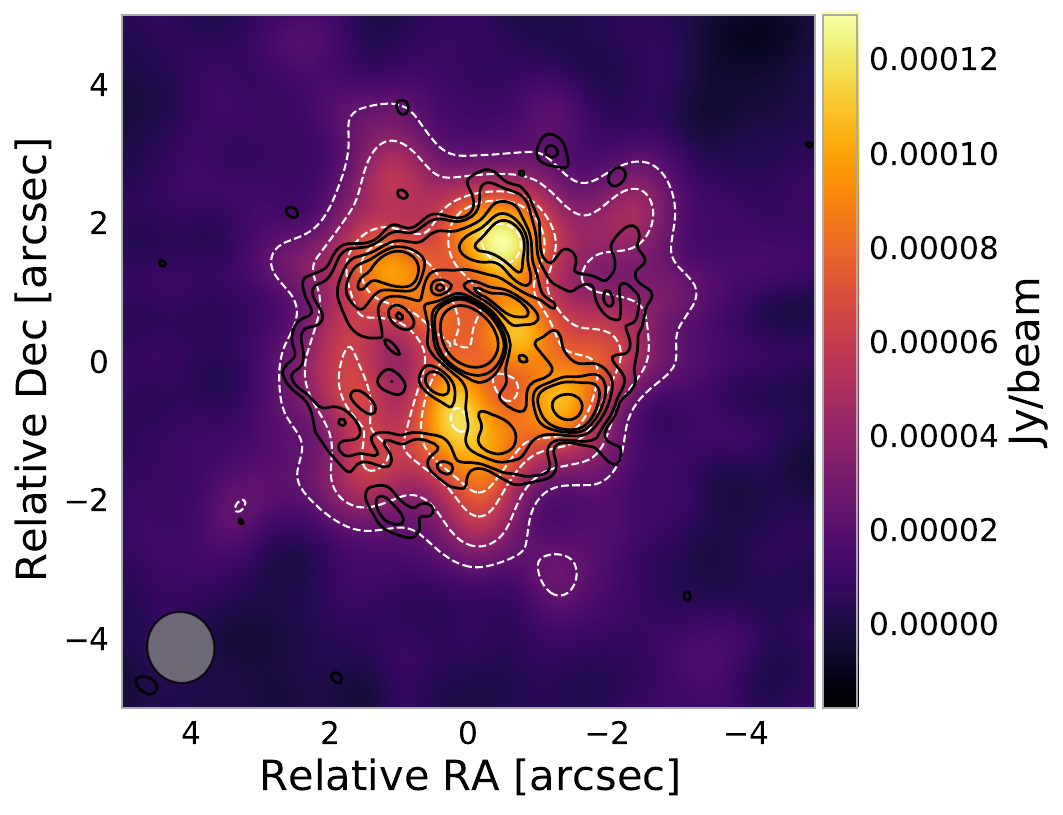} \includegraphics{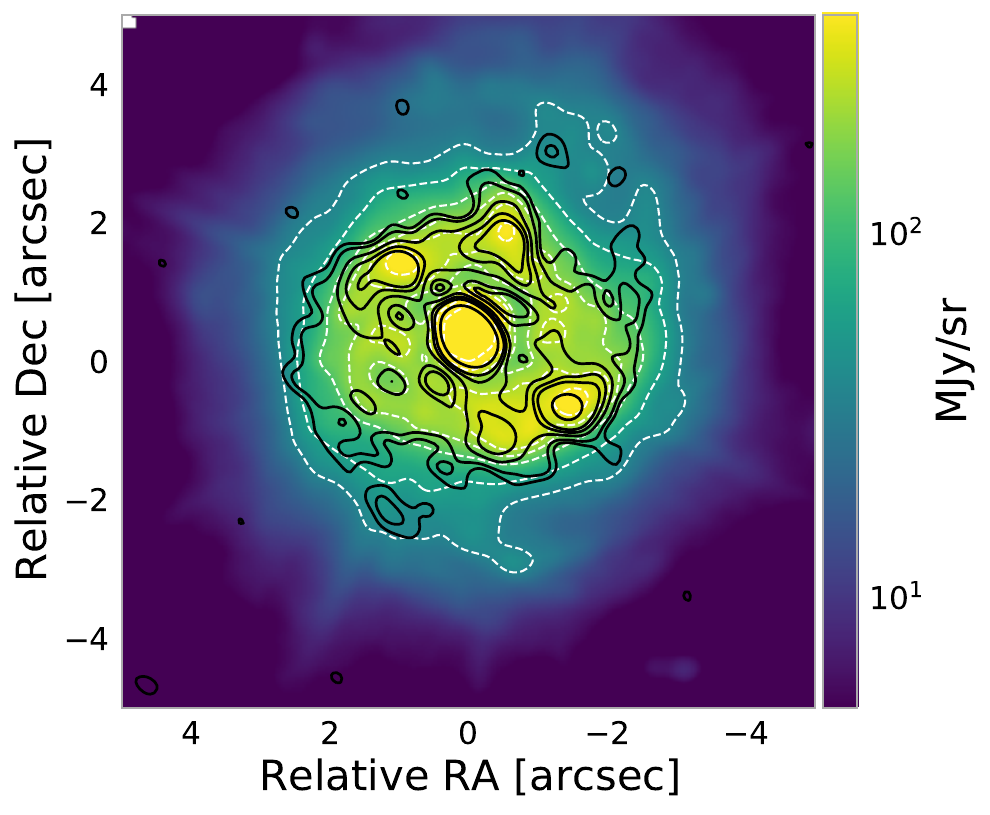}}
 \caption{Left panel: VLA 22 GHz radio continuum map (color scale), with corresponding white contours drawn at (3, 5, 8, 10, 15)$\rm \sigma$, $\rm \sigma = 0.023$ mJy/beam. The clean beam (1.03 arcsec $\times$ 0.97 arcsec, PA=-13 deg) is indicated by the gray ellipse. Right panel: JWST MIRI/F770W  image (color scale), with white contours drawn at (3, 5, 10, 15, 20, 30, 50)$\rm \sigma$, $\rm \sigma = 27$ MJy/sr. The black contours in both panels show the 1.3 mm continuum emission (same as in Figure \ref{fig:alma}, left panel).}
  \label{fig:vla&jwst}
\end{figure*} 

\begin{figure*}[ht]
  \resizebox{\hsize}{!}{ \includegraphics{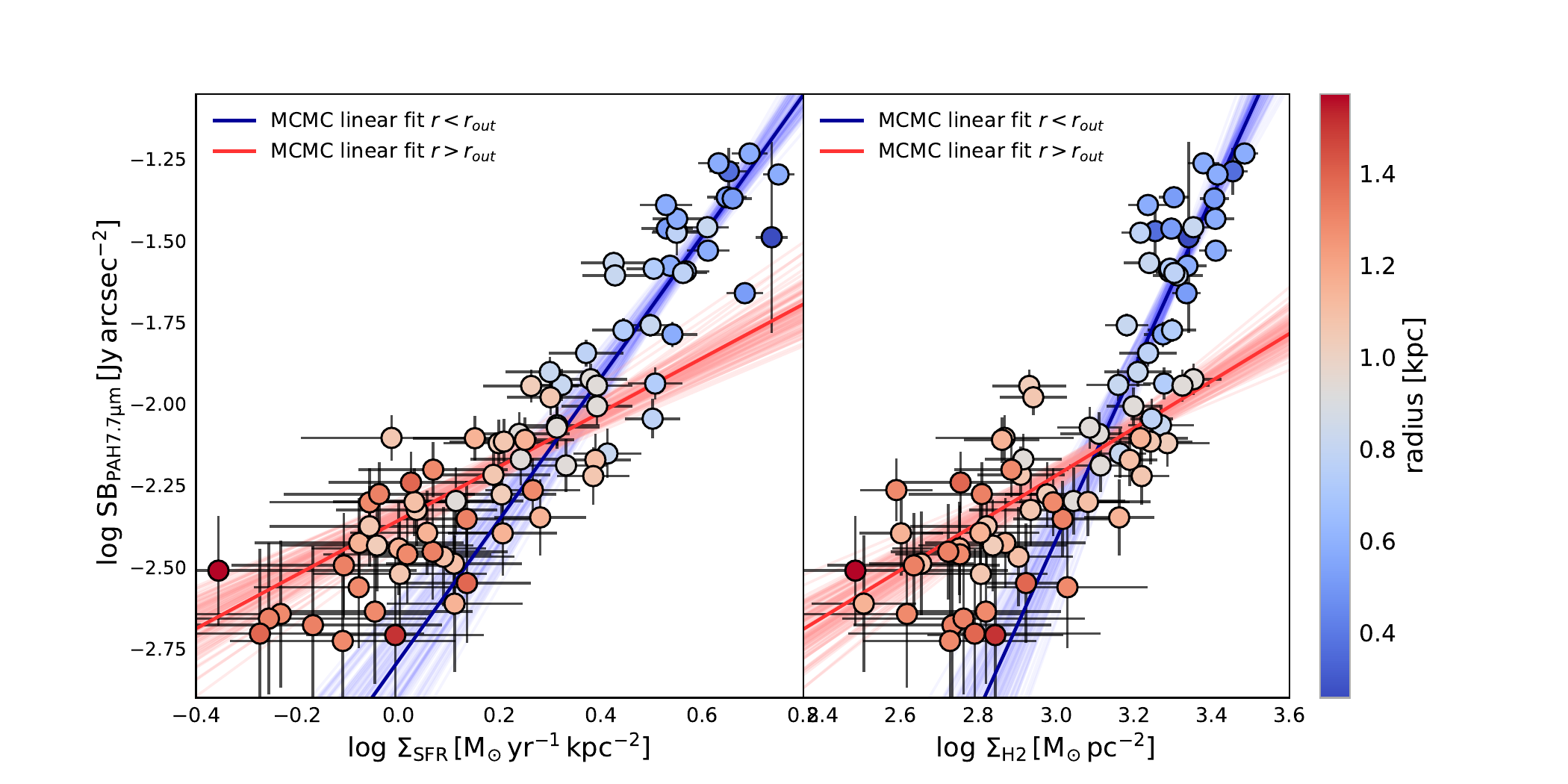}}
 \caption{Surface brightness at 7.7 $\rm \mu m$ from JWST/MIRI imaging versus $\rm \Sigma_{SFR}$ (left panel), and $\rm \Sigma_{H_2}$ (right panel), within the central 10 arcsec $\times$ 10 arcsec region of NGC 7469. The symbols are colored according to their distance from the galaxy center (color bar). The dashed blue (red) lines are the fits to data for $\rm r<r_{SB}$ ($\rm r>r_{SB}$), where $\rm r_{SB}\sim 900 pc$ is the outer SB ring radius \citep{song2021}. 
 }
  \label{fig:F770W}
\end{figure*}

\begin{table*}
     \caption{Derived parameters for the $\rm \Sigma_{SFR}$ - $\rm SB_{PAH}$ and $\rm \Sigma_{H2}$ - $\rm SB_{PAH}$ correlations.}
         \label{tab:corr_coeff}
\centering
\begin{tabular}{l c | c c c c | c c c c }        
\hline\hline         
        & & \multicolumn{4}{c}{$\rm \Sigma_{SFR} - \rm SB_{PAH}$} & \multicolumn{4}{c}{$\rm \Sigma_{H2} - \rm SB_{PAH}$} \\
        \hline
    radius & number data points & $\rho$ & $\tau$ & $n$ & $k$ & $\rho$ & $\tau$ & $n$ & $k$\\ 
    (a)  & (b)  & (c) & (d) & (e) & (f) & (c) & (d) & (e) & (f) \\
    \hline
 $\rm r > 0$ & 82 & 0.91 & 0.75 & 1.77 $\pm$ 0.08 & -2.56 $\pm$ 0.03 & 0.84 & 0.68 & 2.17 $\pm$ 0.11 & -15.48 $\pm$ 0.36 \\
  $\rm r < r_{SB} $ & 31 & 0.76 & 0.56 & 2.16 $\pm$ 0.20 & -2.78 $\pm$ 0.02  & 0.71 & 0.52 & 2.61 $\pm$ 0.25 & -10.25 $\pm$ 0.03 \\
   $\rm r > r_{SB}$ & 51  &0.77 & 0.56 & 0.83 $\pm$ 0.10 & -2.36 $\pm$ 0.03 & 0.66 & 0.47 & 0.72 $\pm$ 0.09 & -4.39 $\pm$ 0.03\\
 \hline
\end{tabular}\\
  \flushleft  \footnotesize{ {\bf Notes.} The columns report (a) the radius of the data points, (b) the number of data points, (c) the Pearson $\rho$ correlation coefficient, (d) the Kendall $\tau$ rank correlation coefficient, (e) the slope $n$ and (f) the zero-point $k$ of the linear fit $y = nx+k$.}
\end{table*}

\begin{figure*}[ht]
  \resizebox{\hsize}{!}{ \includegraphics{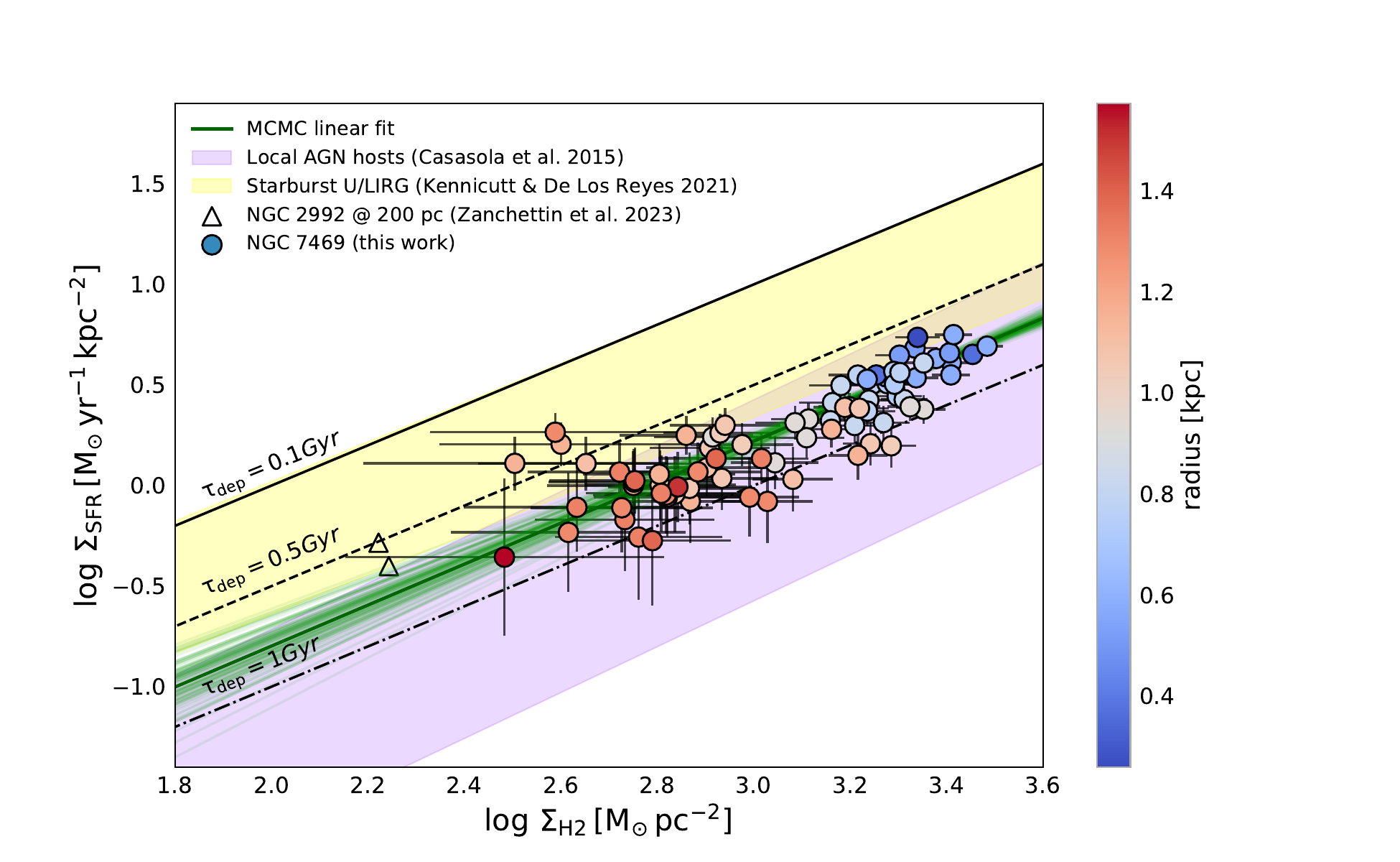}}
 \caption{Spatially resolved star formation law within the central 10 arcsec $\times$ 10 arcsec region of NGC 7469. Circles show the derived quantities in each cell of the grid, coloured according to their distance from the AGN (color bar).
 The dashed green lines are the linear fit to the data $\rm log \Sigma_{SFR} = (1.02 \pm 0.07)log \Sigma_{H2}+(-2.85 \pm 0.02)$, derived by using the python \textit{emcee} library \citep{emcee}.
 Triangles show the derived quantities for NGC 2992 \citep{zanchettin2023}.
 The yellow shaded area shows the best-fit correlation with 0.33 dex scatter found using global measurements for \citep[]{kennicutt2021}: $\rm log \Sigma_{SFR} = (0.98 \pm 0.07)log \Sigma_{H2}+(-2.27 \pm 0.21)$. The purple shaded area shows the range of values found by \citet{casasola2015} for local AGN host galaxies: $\rm log \Sigma_{SFR} = (1.14 \pm 0.01)log \Sigma_{H2}+(-3.41 \pm 0.02)$. Solid, dashed, and dotted black lines represent gas depletion times $\tau _{dep}$ of 0.1, 0.5, and 1 Gyr, respectively. }
  \label{fig:SKplot}
\end{figure*} 

\section{Analysis} \label{sec:analysis}

Within CASA, we modeled and subtracted the central point source emission in the VLA map to remove its contribution from the circumnuclear region. We used the task \texttt{os.system} to model a point source at the AGN position \citep[23:03:15.617 +08:52:26.00 J2000,][]{izumi2020} with a flux of 5.037 mJy at 22 GHz, corresponding to 90\% of the central beam flux. We performed several tests with different point source flux normalization, and checked that with the assumed fraction we avoid artificial over-subtracted regions. The final circumnuclear region flux is corrected by 15 \% and its value is consistent with the central beam flux after point source subtraction.
We used the \texttt{uvsub} task to subtract the point source model from the visibility data, so obtaining the final visibilities. We followed the cleaning procedure described in section \ref{sec:obs} to obtain the final map reported in Figure \ref{fig:vla&jwst}, left panel. 

We used the WebbPSF\footnote{\href{https://webbpsf.readthedocs.io/en/latest/}{https://webbpsf.readthedocs.io}}
tool version 1.2.1 \citep{perrin2014} to simulate the point-spread function (PSF) of the JWST/MIRI F770W filter image, with a pixel size of 0.027 arcsec. We modeled the radial PSF contribution by using circular sectors with a radius equal to the grid cells size, see Section \ref{sec:analysis}, and we derived the radial correction for the F770W filter map.
We corrected both the F770W filter fluxes extracted with the grid and their associated errors for the PSF contribution according to their distance from the center, by following standard error propagation (i.e., convolving the square of the error map with the square of the PSF error). Due to the limitations in the current theoretical model for the JWST PSF, we do not have a robust method for removing the PSF spikes \citep[see][and references therein]{williams2024}. Therefore we applied this conservative method to subtract and model the PSF contribution. We excluded from the analysis the central 1 arcsec ($\sim$330 pc) region where the PSF contribution is dominant. 

We used the continuum-subtracted velocity-integrated CO(2-1) line intensity map to derive the mass surface density of the cold molecular gas, VLA radio map at 22 GHz to estimate the SFR surface density and JWST/MIRI F770W filter map to estimate the PAH emission at 7.7 $\rm \mu m$. Since we analyzed one single PAH feature at 7.7 $\rm \mu m$, we omit the 7.7 $\rm \mu m$ wording from this point onward.
We extract fluxes from a common grid applied to all maps. 
First, we regridded all the maps at the VLA spatial resolution, i.e. the coarser resolution of our data.
The grid is centered at the AGN position and has 1 arcsec wide (i.e. 0.330 kpc) squared cells, equal to regridded maps resolution, 
and a total size of 10 arcsec $\times$ 10 arcsec (i.e. 3.3 kpc $\times$ 3.3 kpc). 
Fluxes are extracted from each cell. 
If the mean flux in the cell is below 3rms per pixel, we consider it as an upper limit. 
For each grid cell, we derived the corresponding cold molecular gas mass ($\rm M_{H2}$) surface density ($\rm \Sigma_{H2}$), SFR surface density ($\rm \Sigma_{SFR}$) and PAH flux density ($\rm \Sigma_{PAH}$). 
The central cell (i.e. the one that contains the AGN) is not used, since it is not possible to determine and disentangle the AGN contribution to the measured flux with good accuracy. Indeed, in this region, we expect both a significant contribution in the JWST image arising from the dust-reprocessed AGN emission from the torus, and a possible contribution to the VLA continuum due to the presence of a core jet-like structure \citep{lodsdale2003,alberdi2006}.
We note that \citet{armus2023} verified that in the total mid-infrared spectrum of NGC 7469 PAH emissions are significantly weaker at the nucleus, due to the AGN heating effect. They reported that the SFR from PAH in the nuclear region, 0.3 arcsec ($\sim$100pc) diameter circular region, is less than 1\% of the total SFR in NGC 7469.

The $\rm \Sigma_{H2}$ in units of $\rm M_{\odot} \, pc^{-2}$ is derived from the velocity-integrated CO(2-1) flux density as: 
\begin{equation}\label{eq:MH2}
     \Sigma_{H2} = \alpha_{CO} \frac{S_{CO(2-1)}}{R_{21}} F_{21} cos~i
 \end{equation}
where $F_{21}$ is the conversion factor from the flux density to the line luminosity $L^\prime$ in K \kms pc$^{2}$ \citep[see][]{solomon2005}, \textit{i} is the galaxy inclination angle, and $\alpha_{\rm CO}$ is the CO-to-$\rm H_2$ conversion factor. We adopt a CO ladder correction factor of $R_{21}$ = 0.7 equal to the mean value found by \citet{casasola2015}, and $\alpha_{\rm CO} = 4.2^{+1.961}_{-1.341} \rm M_{\odot} (K\,km \, s^{-1} \,pc^{2})^{-1} $ derived adopting equation 25 in \citet{accurso2017}, for a stellar mass of $\log(M_*/M_{\odot}) = 10.5 $ \citep{diaz-santos2007}. The $cos~i $ factor is the correction for galaxy inclination, which we assume  \textit{i}=45 deg \citep{davies2004}.

Assuming that most of the radio 22 GHz continuum emission arises from SF, we adopt the radio luminosity at 22 GHz to derive the SFR.
First, we derived the radio luminosity at 1.4 GHz by assuming  an average radio spectrum steepness $\alpha=-0.85$ for the entire galaxy.
We obtained the $\alpha$ parameter by fitting the radio Spectral Energy Distribution (SED) using the integrated radio fluxes reported by NED\footnote{\href{https://ned.ipac.caltech.edu}{https://ned.ipac.caltech.edu}}. The $\alpha$  parameter adopted is consistent with the values used to describe SF regions in normal galaxies and LIRGs \citep[e.g.][]{murphy2012,linden2019,linden2020} and with the values derived by \citet{song2021}.
However, we cannot exclude local variations in the steepness of the radio spectrum that can affect the final SFR values. \citet{song2021} derived the spectral index measured from 15 to 33 GHz, finding values in between $\sim$-0.7 and $\sim$-0.9 in the NGC 7469 nuclear ring. These variations translate into corrections to the SFR up to 10$\%$ which will be taken into account further in the discussion.
We then obtained the SFR adopting the correlation reported by \citet{shao2018} in section 3.2, and the corresponding star formation surface density, $\rm \Sigma_{SFR}$, by correcting for galaxy inclination, i.e. multiplying by the $cos~i $ factor as in equation \ref{eq:MH2}.

The MIRI F770W allows us to sample the 7.7 $\rm \mu m$ PAH emission complex in NGC7469. Although the band is dominated by the PAH emission feature, it may also include continuum emission due to hot dust. \citet{whitcomb2023} showed that the Spitzer/IRS spectra of the star-forming regions in 75 nearby galaxies (SINGS sample) may have up to 10\% continuum contamination in the F770W filter. Therefore, in a conservative approach, we assumed that 90\% of the flux is attributed to PAH \citep{chastenet2023}.

\section{Discussion and Conclusion}
\label{sec:discussion}

\subsection{Morphology of the SB ring}

We estimate the global SFR integrated over the central 3.3 kpc $\times$ 3.3 kpc into SFR = $11.5 \pm 0.3$ \sfr, 
following the prescription of \citet{shao2018}, see Section \ref{sec:analysis}. This estimate does not include the contribution from the central 330 pc $\times$ 330 pc cell.
The uncertainty accounts for a realistic estimate of the systematic scatter in the 1.4 GHz-based method \citep[$\sim$0.3 dex according to][]{shao2018}. 
Our SFR estimate is consistent with the SFR = $21 \pm 2.1$ \sfr\ by \citet{song2021} from the 3-to-33 GHz radio continuum, and with the range of SFR = 10-30 \sfr\ reported by \citet{lai2022} using optical emission lines, namely the $\rm Pf \alpha$ recombination line, [Ne II] and [Ne III].
We found that the majority, approximately 60\%, of the SF occurs in the SB ring, consistent with \citet{song2021}.

We computed the molecular gas mass over the central 3.3 kpc $\times$ 3.3 kpc $\rm M(H2) = (6.4 \pm 2.0) \times 10^9 M_{\sun}$, where the error is dominated by the uncertainty on $\alpha_{\rm CO}$ \citep[0.165 dex,][]{accurso2017}. This estimate is consistent within 2 sigma with the values reported in the literature using different CO transitions \citep{davies2004, izumi2020,nguyen2021,zhang2023}. We note that these works focus on the inner $\sim$1 kpc region of NGC 7469, i.e. the SB ring.
We computed that $\sim 50\%$ of the total cold molecular gas mass is contained in the SB ring, $\rm M(H2)_{SB} = (3.3 \pm 1.0) \times 10^9 M_{\sun}$.
The central concentration of gas and SFR is in line with those observed in local LIRGs \citep{diazsantos2010,diazsantos2011}.
Adopting our estimate of $\rm M(H2)$ and SFR, we derived the characteristic gas depletion timescale, $\rm \tau_{dep}$ = $\rm M(H_2)/SFR$. We obtained $\rm \tau_{dep} = (0.56 \pm 0.18)~Gyr$ for the central 3.3 kpc $\times$ 3.3 kpc region, and a shorter $\rm \tau_{dep} = (0.47 \pm 0.15)~Gyr$ for the SB ring. 
Considering the errors associated with the values, the depletion times are consistent within one sigma.

Figure \ref{fig:alma} shows the distribution of the 1.3 mm continuum and the integrated CO(2-1) flux map (Moment 0 map). In the continuum map, we identify five representative positions (A–E), that are detected as clumps with SNR$>15$. Clumps from A to D are named as in \citet{izumi2020}, who first identified the clumps using the ALMA 860 $\rm \mu m$ continuum map. In addition, we identify clump E (23:03:15.60 +08:52:24.56, J2000) where we detect 1.3 mm emission with a statistical significance of $\sim$18$\rm \sigma$.
The 1.3 mm central peak position, labeled as clump A, is consistent with the AGN position from VLA 3.5 cm and ALMA 860 $\rm \mu m$ continuum emission \citep{condon1991,orientiprieto2010, izumi2020}.
Due to the PSF smearing of the central AGN emission in MIRI 7.7 $\rm \mu m$, we excluded from the following analysis the central cell, which contains also the CND \citep{davies2004,izumi2015}.
Figure \ref{fig:vla&jwst} shows the VLA 22 GHz emission map (left panel) and the JWST/MIRI PAH emission map (right panel), both with 1.3 mm emission contours overlaid. 
The 1.3 continuum map traces the dust distribution across the SB ring, which shows a morphology that is remarkably similar to those traced by PAHs and radio continuum at 22 GHz. The three tracers reproduce the morphology of the SB ring with good agreement.
In particular, clumps B, C, and D are detected with similar morphology and size in the dust continuum, CO(2-1), 22 GHz, and the PAH emission. 
The overall clumps distribution appears to be consistent also with the 860 $\rm \mu m$ continuum \citep{izumi2020}, suggesting that the 1.3 mm emission is mainly tracing thermal dust emission heated by young stars in the SB ring.
While the dusty clumps B, C, D, and E all show prominent counterparts also in the CO map (Figure \ref{fig:vla&jwst}), the dusty clump E exhibits no evident counterpart in the radio or mid-IR maps.
Interestingly, the 22 GHz radio continuum map presents an emission peak with a significance of 15$\rm \sigma$ (23:03:15.63 +08:52:24.98 J2000), offset by about 0.5 arcsec with respect to E along the north-east direction (white and black contours in the left panel of Figure \ref{fig:F770W}). We exclude a contribution from the unresolved radio jet at this location, as a core jet-like structure was only detected at the nucleus \citep{lodsdale2003,alberdi2006}. Indeed, the local enhancement in the radio emission may be attributed to a gas inflow associated to the SB ring and to the two-arm spiral structure, suggested by the high velocity dispersion of the cold molecular gas at this location \citep{nguyen2021}. 
Clumps B and D are located at the endpoints of a bar-like feature identified by \citet{davies2004,fathi2015, izumi2015}. Although the bar structure is detected at a significance level of 12$\rm \sigma$ in the CO(2-1) map, it remains undetected down to 3$\rm \sigma$ in the radio continuum and in the 7.7 $\rm \mu m$ maps, despite the high angular resolution of JWST/MIRI.

\subsection{PAH emission}

In the following, we investigate the correlation between $\rm \Sigma_{SFR}$,  $\rm \Sigma_{H_2}$ and PAH surface brightness, $\rm SB_{PAH}$.
Figure \ref{fig:F770W} shows $\rm SB_{PAH}$ versus $\rm \Sigma_{SFR}$ (left panel), and $\rm \Sigma_{H_2}$ (right panel), within the region of NGC 7469 sampled by the grid.
The PAH surface brightness correlates both with the SFR and the cold molecular gas surface densities. We derived the Pearson $\rho$ and the Kendall $\tau$ rank correlation coefficients (Table \ref{tab:corr_coeff}), finding a higher correlation of $\rm SB_{PAH}$ with $\rm \Sigma_{SFR}$ than with $\rm \Sigma_{H2}$, both for the SB ring and for external regions. 
Although the signal-to-noise ratio (SNR) of the radio map is lower than that of the CO map (Section \ref{sec:obs}),
the correlation between $\rm SB_{PAH}$ and $\rm \Sigma_{SFR}$ is stronger and narrower than that with $\Sigma_{H2}$. 
This is likely due to PAH emission being more affected by the ionizing radiation field of young stars rather than by the availability of molecular gas \citep[see][and references therein]{jensen2017, belfiore2023}. 

We also observe a steepening of both relations moving towards the center of the galaxy, as highlighted in Figure \ref{fig:F770W} by the color gradient of the symbols. 
To quantify this steepening, we divided the data into two subsamples: an inner region within the SB ring ($\rm r < r_{SB}$) and an outer region with $\rm r > r_{SB}$, where $\rm r_{SB} = 900~pc$. This value has been measured by \citep{song2021} using the 3 to 30 GHz continuum and is consistent with our maps. 
We fit a linear relation $y = nx+k$ to each subsample using  Python package \textit{emcee} \citep{emcee}, and assuming a constant error on $\rm \Sigma_{H2}$ and $\rm \Sigma_{SFR}$, equal to the rms noise of each {regridded map. The uncertainty associated to $\rm SB_{PAH}$ is obtained by convolving the rms noise of the regridded MIRI map with the uncertainty due to the PSF contribution (Section \ref{sec:analysis}).
Blue and red solid lines in Figure \ref{fig:F770W} show the fits for $\rm r<r_{SB}$ and for $\rm r>r_{SB}$, respectively.  
For both  $\rm \Sigma_{SFR}$ and $\rm \Sigma_{H2}$, the SB ring shows a steeper slope with respect to $\rm r>r_{SB}$ (Table \ref{tab:corr_coeff}). In particular, the slope at the SB ring is $n=( 2.16 \pm 0.20)$ 
whereas at $\rm r>r_{SB}$ is $n=(0.83 \pm 0.10)$. A similar behavior is seen for the $\rm \Sigma_{H2}$ correlation.  
The steepening of these relations may be attributed to an enhancement of recent SF activity in the SB ring structure, as already observed using the 11.3\micron\ PAH feature by \citet{jensen2017}. 
This is consistent with the presence of a young population of ionizing stars in the SB ring, as reported by \citet{diaz-santos2007}. 
\citet{jensen2017} measured a negligible contribution of AGN-heated dust to the continuum flux at 11\micron\ above 200 pc.
Therefore, we exclude that this steepening is due to an increased contribution of AGN-heated dust to the continuum emission measured in the JWST F770W filter, at the location of the SB ring.
Regarding a possible AGN contribution to
PAH destruction, we do not expect a significant effect given the modest luminosity of our target \citep[$\rm L_{bol,AGN}=3.5\times 10^{44}$ \ergs\ ,][]{gruppioni2016}.  

 Still local variations in gas properties, including composition, turbulence, and metallicity, significantly influence the CO-to-H2 conversion factor \citep{leroy2011,narayan2012,papadopoulos2012,bolatto2013,sandstrom2013,teng2024}.
Oxygen abundances compatible with solar metallicity were reported for NGC 7469 SB ring and outer region by \citet{cazzoli2020} and \citet{zamoradiaz2023} using optical nebular lines. However, metallicity determination is challenging at all spatial scales, particularly in AGN-affected nuclear regions where it remains unknown.
Considering a CO-to-H2 conversion factor accounting for both solar-like metallicity and gas excitation, as prescribed by \citet{narayan2012}, we found that the change in slope between SB ring and outer disk remains appreciable. 
A significantly decreasing metallicity gradient toward the nucleus could still mitigate the slope change in the $\mathrm{SB_{PAH}-\Sigma_{H2}}$ relation. However, there are no evidence for such trend.
Additionally, several factors, including grain size distribution, temperature, ionization field, metallicity, and dust amount can affect PAH luminosity \citep[e.g.][]{chastenet2023,whitcomb2024}. The 7.7 \micron\ complex is associated with larger, positively charged ions \citep{galliano2008,draine2021,rigopoulou2021,maragkoudakis2022}, and its emission can increase around young stars due to a stronger radiation field. However, investigating PAH properties such as grain size distribution and ionization fraction involves examining the ratios between various MIR PAH features \citep{lai2023, bohn2023}.
Finally, spatially resolved studies in nearby galaxies have reported a steepening of the radio spectral index with increasing distance from the center \citep[e.g.][]{paladino2009,basu2015, roy2021}, with additional local variations in regions of active SF \citep{westcott2018}. 
We therefore cannot exclude that the superposition of these effects could cause the observed slope change in the $\rm log SB_{PAH}-log \Sigma_{SFR}$ relation.


\subsection{The star formation law}
 
Figure \ref{fig:SKplot} shows the spatially resolved star formation law on a $\sim 300$ pc scale for NGC 7469. The plot also includes NGC 2992 measurements obtained with a similar method on $\sim 200$ pc scales \citep{zanchettin2023}, and the relations derived for compilation of local AGN \citep{casasola2015}, and ultraluminous/luminous infrared galaxies \citep[U/LIRGs, $ \rm  L_{IR} > 10^{12} \, L_{\odot}$, and $\rm > 10^{11} \, L_{\odot}$, respectively,][]{kennicutt2021}. 
The purple shaded area shows the K-S relation for local AGN hosts drawn from the NUclei of GAlaxies (NUGA) survey, characterized by SFR within $\rm -0.9 < log~SFR / M_{\odot} yr^{-1} < 0.8$ and cold molecular gas masses in the range $\rm 8.5 < log M(H_2) / M_{\odot} yr^{-1} < 9.6$ \citep[see][and references therein]{casasola2015}.
The yellow shaded are shows the star formation law found for dusty starbursts U/LIRGs, with $\rm 1.1 < log~SFR / M_{\odot} yr^{-1} < 1.8$, and cold molecular gas masses $\rm 9.4 < log M(H_2)/ M_{\odot} yr^{-1} < 10.2$ \citep{kennicutt2021}. 
We adopted the same fitting method as described above, taking into account the uncertainties both on $\rm \Sigma_{SFR}$ and on $\rm \Sigma_{H2}$, and the best fit linear relation for all data points is $\rm log \Sigma_{SFR} = (1.02 \pm 0.07) log \Sigma_{H2}+(-2.85 \pm 0.02)$ (green solid line in Figure \ref{fig:SKplot}). The slope of this relation is consistent with that obtained by fitting the SB ring alone, while we find a flatter slope for the fit of the outer region ($n=0.5 \pm 0.1$). As the latter fit is dominated by data points with low SNR, we consider the global $\rm log \Sigma_{SFR}-log \Sigma_{H2}$ relation as best-fit for NGC 7469.

The star formation relation derived for the NUGA AGN (purple shaded area) exhibits a slightly steeper slope ($n = 1.14 \pm 0.01$) and a $\sim$0.6 dex offset towards lower SFR surface densities with respect to NGC 7469.
\citet{kennicutt2021} found that starburst galaxies and U/LIRGs are fitted by a linear relation with slope $n = (0.98 \pm 0.07)$, consistent with the one derived for NGC 7469 in this work, and a $\sim$0.6 dex offset in the zero-point towards higher SFR surface densities (yellow shaded area).
Considering the associated errors, the slopes of the relations reported by \citet{kennicutt2021} and \citet{casasola2015} are roughly consistent within three sigma.
Figure \ref{fig:SKplot} indicates that the individual regions of NGC 7469 occupy an intermediate area of the $\rm log \Sigma_{SFR}-log \Sigma_{H2}$ plane, corresponding to an intermediate star formation regime between local AGNs and U/LIRGs.
The observed scatter in the data points can reflect local variations in the physical conditions of the ISM from which stars form, such as the morphology and concentration of the gas, and any differences in gas excitation. Additionally, local fluctuations in physical quantities affect the conversion (e.g., CO-to-H2 conversion factor, radio steepness) from observables to physical quantities.
In Figure \ref{fig:SKplot}, we report the characteristic gas depletion timescales: 
NGC 7469 data points show an average $\rm \tau_{dep, mean}$ = ($0.66 \pm 0.02$) Gyr, where the uncertainty is the standard error of the mean.
The measurements of the circumnuclear region of NGC 2992 \citep[black triangles,][]{zanchettin2023}, derived with a spatial resolution of 200 pc, populate regions of the star formation law chart characterized by both relatively lower $\rm log \Sigma_{SFR}$ and $\rm log \Sigma_{H2}$, with respect to NGC 7469.
The difference observed between NGC 2992 and our target may reflect the fact that individual galaxies may show different star formation laws, as highlighted by \citet{casasola2022} in their work on a set of nearby spiral DustPedia\footnote{\href{http://dustpedia.astro.noa.gr/}{http://dustpedia.astro.noa.gr/}} galaxies.
However, NGC 2992 has an average $\rm \tau_{dep, mean}$ = ($0.38 \pm 0.07$) Gyr, consistent with that measured in our target.

\section{Acknowledgments}.
\begin{acknowledgments}
This work is based in part on observations made with the NASA/ESA/CSA James Webb Space Telescope. The data were obtained from the Mikulski Archive for Space Telescopes at the Space Telescope Science Institute, which is operated by the Association of Universities for Research in Astronomy, Inc., under NASA contract NAS 5-03127 for JWST. These observations are associated with program No. 1328. The {\it JWST} data used in this paper can be found in MAST: \dataset[10.17909/3x02-0q74]{http://dx.doi.org/10.17909/3x02-0q74}.
This paper makes use of the following ALMA data: ADS/JAO.ALMA\#2017.1.00078.S ALMA is a partnership of ESO (representing its member states), NSF (USA) and NINS (Japan), together with NRC (Canada), MOST and ASIAA (Taiwan), and KASI (Republic of Korea), in cooperation with the Republic of Chile. The Joint ALMA Observatory is operated by ESO, AUI/NRAO and NAOJ.
Based on observations No. 20A-158 with the National Radio Astronomy Observatory which is a facility of the National Science Foundation operated under cooperative agreement by Associated Universities, Inc.
We acknowledge financial support from PRIN MIUR contract 2017PH3WAT, and PRIN MAIN STREAM INAF "Black hole winds and the baryon cycle".
We acknowledge support from the INAF Mini Grant GO/GTO "SHORES: Serendipitous H-ATLAS fields Observations of Radio Extragalactic Sources".
M.B. acknowledges support from INAF project 1.05.12.04.01 - MINI-GRANTS di RSN1 "Mini-feedback" and FVG LR 2/2011 project D55-microgrants23 "Hyper-gal".

\end{acknowledgments}

%

\vspace{5mm}
\facilities{JWST(MIRI), ALMA, VLA}


\software{astropy \citep{2013A&A...558A..33A,2018AJ....156..123A}, emcee \citep{emcee}, CASA \citep{mcmullin}, WebbPSF \citep{perrin2014}}





\bibliography{sample631}{}
\bibliographystyle{aasjournal}



\end{document}